# Network Visualization of ChatGPT Research: a study based on term and keyword co-occurrence network analysis

Deep Kumar Kirtania

Librarian, Bankura Sammilani College

deepkrlis@gmail.com

**Abstract:** The main objective of this paper is to identify the major research areas of ChatGPT through term and keyword co-occurrence network mapping techniques. For conducting the present study, total of 577 publications were retrieved from the Lens database for the network visualization. The findings of the study showed that "chatgpt" occurrence in maximum number of times followed by its related terms such as artificial intelligence, large language model, gpt, study etc. This study will be helpful to library and information science as well as computer or information technology professionals.

**Introduction:** Visualizing bibliometric networks refers to the use of visualization tools and techniques to represent and analyze the relationships between different academic publications based on bibliometric data, such as citations, co-authorship, and co-citations (Perianes-Rodriguez, Waltman & Van Eck, 2016; Najmi et al., 2017; Liao et al., 2018; Jiang, Ritchie & Benckendorff, 2019). Visualization techniques used for bibliometric networks include graphs, maps, and network diagrams, which can provide insights into the structure, evolution, and impact of research areas, institutions, or individual scholars (Lee & Chen, 2012; Pradhan, 2017; Aparicio, Iturralde & Maseda, 2019).These visualizations can help researchers identify key authors, important publications, and emerging trends in their fields, and can also be useful for policymakers and funding agencies looking to evaluate the impact and influence of academic research (Ellegaard & Wallin, 2015; Reed et al., 2021). Visualizing bibliometric networks is a powerful tool for understanding the complex relationships between academic publications and can provide valuable insights into the dynamics of scholarly communication and collaboration (Zhang & Eichmann-Kalwara, 2019; Shasha et al., 2020). Term and keyword co-occurrence refers to the phenomenon where certain terms or keywords appear together frequently in a text corpus, indicating a relationship or association between them (Scott & Tribble, 2006; Hunston, 2007; Santoni & Pourabbas, 2016; Lozano et al, 2019). Term and keyword co-occurrence analysis can be a powerful tool for identifying meaningful patterns in text data and gaining insights into the relationships between different concepts and keywords (Bondi, 2010; Cech, 2017; Radhakrishnan, Erbis, Isaacs & Kamarthi, 2017; Weerasekara et al, 2022). These network visualization tools can be helpful in analyzing the usage and impact of ChatGPT's keywords in academic publications. Recently two research papers have been carried out on ChatGPT and bibliometrics (Kirtania, 2023; Farhat, Sohail & Madsen, 2023) and here ChatGPT is discussed with advantages, limitations and trustworthiness. Visualizing this type of networks for analyzing the impact of ChatGPT's research and its keywords in the academic community, and can help researchers gain valuable insights into the broader research landscape. The present paper makes a small attempt to do this.

**Objective:** The main objective of this study is to identify the major research areas of ChatGPT via term and keyword co-occurrence network mapping techniques.

**Methodology:** The Lens (https://www.lens.org/) database was searched with scholarly works under ChatGPT and 577 research outputs were considered for the present study. The Lens or Patent Lens is an online patent and scholarly literature search database and analytics tool which providing access to a global corpus of scholarly literature metadata with citation indexing. Those 577 publications were downloaded in excel format. Later those excel files were used in network analysis through VOSviewer (https://www.vosviewer.com/) software and keywords were extracted from there. Finally, the data visualization analysis has been done with the help of VOSviewer software and Word Cloud of major research area completed with the help of Lens database.

**Limitation:** There are not many publications on Chat GPT as it is a very recent topic. This is a limitation of the current study because, the more publications on this type of work, the more valid the research will be.

# Visualization of the Results

**Fig 1: Network Visualization of term co-occurrence of title and abstract**

**Fig 2: Overlay Visualization of term co-occurrence of title and abstract**

**Fig 3: Network Visualization of term co-occurrence of abstract**

**Fig 4: Network Visualization of Keyword co-occurrence of all keywords**

**Fig 5: Network Visualization of Keyword co-occurrence of author keywords**

**Fig 5: Word cloud of Top fields of study (https://www.lens.org/)**

**Major Findings of the Study**

The major findings of this study are:

I. 6908 terms were obtained from 577 publications by extracting the term co-occurrence of title and abstract filed. 50 most occurrencewords out of 6908 terms are given in Appendix 1 where terms, their occurrences time, relevance score and percentage are given. From that table it can be seen that the term chatgpt score the highest position with 1276 occurrence, followed by question (215), study (208), model (188) etc. The strong relationship between these terms is easily seen from the Network Visualization image (Fig 1, 2 & 3).

II. Keyword Co-occurrence mapping (Fig 4) of all keywords shows that artificial intelligence is at the first place followed by chatgpt, humans, machine learning, ethics, publishing etc.

III. Keyword Co-occurrence mapping (Fig 5) of author keywords shows that chatgpt, artificial intelligence and machine learning are the most use term followed by ethics, publishing, ai, chatbot, medical education etc.

IV. Top research areas of ChatGPT shows that most papers are published under the field of computer science (209), followed by artificial intelligence (123) and psychology (115) respectively. In addition, many other topics are already being worked on with ChatGPT, as it can be easily seen from the word cloud (Fig 5) provided by the Lens database. From the subject trends, it is easy to say that the application of ChatGPT in various topics is currently being worked by the researcher globally and more research will be done on this topic in the future.

**Conclusion:** Network visualization, co-occurrence network analysis has become a very important subject in the current context. With the help of this we can understand the relationship of various entities and we can easily analyze the research based relationship between them. In the present paper, a visualization network analysis of all the publications published on ChatGPT has been carried out. The present paper shows that keywords related to ChatGPT are the most used and these words have the highest occurrence and relevance score. ChatGPT is a completely recent topic and there is still a lot of research to be done on this topic. It is easy to expect that there will be much more work on Chat GPT in the future, and more network analysis of that work. That is, it can be easily said that more research work can be done on this topic in the future.

**References:**

Aparicio, G., Iturralde, T., & Maseda, A. (2019). Conceptual structure and perspectives on entrepreneurship education research: A bibliometric review. *European research on management and business economics*, *25*(3), 105-113.

Bondi, M. (2010). Perspectives on keywords and keyness. *Keyness in texts*, 1-20.

Cech, F. (2017). Exploring emerging topics in social informatics: an online real-time tool for keyword co-occurrence analysis. In *Social Informatics: 9th International Conference, SocInfo 2017, Oxford, UK, September 13-15, 2017, Proceedings, Part II 9* (pp. 527-536). Springer International Publishing.

Ellegaard, O., & Wallin, J. A. (2015). The bibliometric analysis of scholarly production: How great is the impact?. *Scientometrics*, *105*, 1809-1831.

Farhat, F., Sohail, S. S., & Madsen, D. Ø. (2023) How Trustworthy is ChatGPT? The Case of Bibliometric Analyses. Preprints.org https://doi.org/10.20944/preprints202303.0479.v1.

Hunston, S. (2007). Semantic prosody revisited. *International journal of corpus linguistics*, *12*(2), 249-268.

Jiang, Y., Ritchie, B. W., & Benckendorff, P. (2019). Bibliometric visualisation: An application in tourism crisis and disaster management research. *Current Issues in Tourism*, *22*(16), 1925-1957.

Kirtania, D. K. (2023). ChatGPT as a Tool for Bibliometrics Analysis: Interview with ChatGPT. *Available at SSRN* https://papers.ssrn.com/sol3/papers.cfm?abstract_id=4391794

Lee, M. R., & Chen, T. T. (2012). Revealing research themes and trends in knowledge management: From 1995 to 2010. *Knowledge-Based Systems*, *28*, 47-58.

Liao, H., Tang, M., Luo, L., Li, C., Chiclana, F., & Zeng, X. J. (2018). A bibliometric analysis and visualization of medical big data research. *Sustainability*, *10*(1), 166.

Lozano, S., Calzada-Infante, L., Adenso-Díaz, B., & García, S. (2019). Complex network analysis of keywords co-occurrence in the recent efficiency analysis literature. *Scientometrics*, *120*, 609-629.

Najmi, A., Rashidi, T. H., Abbasi, A., & Travis Waller, S. (2017). Reviewing the transport domain: An evolutionary bibliometrics and network analysis. *Scientometrics*, *110*, 843-865.

Perianes-Rodriguez, A., Waltman, L., & Van Eck, N. J. (2016). Constructing bibliometric networks: A comparison between full and fractional counting. *Journal of informetrics*, *10*(4), 1178-1195.

Pradhan, P. (2017). Science mapping and visualization tools used in bibliometric & scientometric studies: An overview. https://ir.inflibnet.ac.in/bitstream/1944/2132/1/INFLIBNET%20NEWSLETTER%20Vol.23%20No.%204%20(October-%20December%202016).pdf

Radhakrishnan, S., Erbis, S., Isaacs, J. A., & Kamarthi, S. (2017). Novel keyword co-occurrence network-based methods to foster systematic reviews of scientific literature. *PloS one*, *12*(3), e0172778.

Reed, M. S., Ferre, M., Martin-Ortega, J., Blanche, R., Lawford-Rolfe, R., Dallimer, M., & Holden, J. (2021). Evaluating impact from research: A methodological framework. *Research Policy*, *50*(4), 104147.

Santoni, D., & Pourabbas, E. (2016). Automatic detection of words associations in texts based on joint distribution of words occurrences. *Computational Intelligence*, *32*(4), 535-560.

Scott, M., & Tribble, C. (2006). *Textual patterns: Key words and corpus analysis in language education* (Vol. 22). John Benjamins Publishing.

Shasha, Z. T., Geng, Y., Sun, H. P., Musakwa, W., & Sun, L. (2020). Past, current, and future perspectives on eco-tourism: A bibliometric review between 2001 and 2018. *Environmental Science and Pollution Research*, *27*, 23514-23528.

Weerasekara, S., Lu, Z., Ozek, B., Isaacs, J., & Kamarthi, S. (2022). Trends in Adopting Industry 4.0 for Asset Life Cycle Management for Sustainability: A Keyword Co-Occurrence Network Review and Analysis. *Sustainability*, *14*(19), 12233.

Zhang, L., & Eichmann-Kalwara, N. (2019). Mapping the scholarly literature found in Scopus on "research data management": A bibliometric and data visualization approach. *Journal of Librarianship and Scholarly Communication*, *7*(1). https://doi.org/10.7710/2162-3309.2266

**Appendix 1: Top 50 term co-occurrence of title and abstract**

| term | occurrences | relevance score | Percentage |
|---|---|---|---|
| chatgpt | 1276 | 0.0381 | 18.47 |
| question | 215 | 0.2896 | 3.11 |
| study | 208 | 0.1293 | 3.01 |
| model | 188 | 0.1187 | 2.72 |
| use | 137 | 0.0585 | 1.98 |
| response | 128 | 0.3439 | 1.85 |
| artificial intelligence | 124 | 0.1263 | 1.80 |
| large language model | 116 | 0.0956 | 1.68 |
| tool | 116 | 0.1185 | 1.68 |
| text | 115 | 0.1253 | 1.66 |
| gpt | 111 | 0.4569 | 1.61 |
| research | 110 | 0.2952 | 1.59 |
| technology | 110 | 0.3435 | 1.59 |
| performance | 108 | 0.2939 | 1.56 |
| chatbot | 103 | 0.4858 | 1.49 |
| paper | 100 | 0.1604 | 1.45 |
| education | 97 | 0.6251 | 1.40 |
| llm | 88 | 0.1658 | 1.27 |
| application | 87 | 0.1891 | 1.26 |
| task | 85 | 0.1249 | 1.23 |
| information | 83 | 0.1857 | 1.20 |
| system | 83 | 0.5392 | 1.20 |
| review | 78 | 0.2389 | 1.13 |
| student | 77 | 0.1696 | 1.11 |
| data | 75 | 0.2729 | 1.09 |
| openai | 72 | 0.0973 | 1.04 |
| challenge | 71 | 0.1577 | 1.03 |
| content | 69 | 0.5895 | 1.00 |
| field | 67 | 0.25 | 0.97 |
| role | 67 | 0.2091 | 0.97 |
| article | 66 | 0.2202 | 0.96 |
| accuracy | 63 | 0.529 | 0.91 |
| capability | 62 | 0.1158 | 0.90 |
| language model | 61 | 0.0713 | 0.88 |
| user | 61 | 0.4332 | 0.88 |
| analysis | 60 | 0.0793 | 0.87 |
| limitation | 60 | 0.122 | 0.87 |
| answer | 59 | 0.2812 | 0.85 |
| implication | 59 | 0.1055 | 0.85 |
| approach | 58 | 0.3713 | 0.84 |

| ability   | 57 | 0.0927 | 0.83 |
| --------- | -- | ------ | ---- |
| patient   | 57 | 0.6724 | 0.83 |
| issue     | 54 | 0.3323 | 0.78 |
| knowledge | 54 | 0.1308 | 0.78 |
| science   | 54 | 0.403  | 0.78 |
| dataset   | 53 | 0.2038 | 0.77 |
| context   | 52 | 0.2209 | 0.75 |
| author    | 51 | 0.3966 | 0.74 |
| case      | 51 | 0.4292 | 0.74 |
| topic     | 51 | 0.1271 | 0.74 |